\documentclass[conference]{IEEEtran}

\usepackage[utf8]{inputenc}
\usepackage[textsize=footnotesize]{todonotes}
\usepackage{tikz}
\usepackage{array}
\usepackage{amssymb}
\usepackage{graphicx}
\usepackage{multirow}
\usepackage{subfigure}
\usepackage{pgfplots, tikz}
\usepackage{pgfplotstable}
\usepackage{amsmath}
\usepackage{relsize}
\usepackage{url}

\newcolumntype{L}[1]{>{\raggedright\let\newline\\\arraybackslash\hspace{0pt}}m{#1}}
\newcolumntype{C}[1]{>{\centering\let\newline\\\arraybackslash\hspace{0pt}}m{#1}}
\newcolumntype{R}[1]{>{\raggedleft\let\newline\\\arraybackslash\hspace{0pt}}m{#1}}

\begin{document}

\title{Improving Blind Steganalysis in Spatial Domain using a Criterion to Choose the Appropriate Steganalyzer between CNN and SRM+EC}



\newcommand{\RC}[2][inline]{%
  \todo[color=blue!10,#1]{\sffamily\textbf{RC:} #2}\xspace}
\newcommand{\MS}[2][inline]{%
  \todo[color=green!10,#1]{\sffamily\textbf{MS:} #2}\xspace}
\newcommand{\NS}[2][inline]{%
  \todo[color=orange!10,#1]{\sffamily\textbf{NS:} #2}\xspace}

\author{
\IEEEauthorblockN{Jean-Francois Couchot\IEEEauthorrefmark{1}, Rapha\"el Couturier\IEEEauthorrefmark{2}, Michel Salomon\IEEEauthorrefmark{3}}
\IEEEauthorblockA{FEMTO-ST Institute, UMR 6174 CNRS - Univ. Bourgogne Franche-Comt\'e (UBFC), Belfort, France\\
DISC (department of computer science and complex systems) - AND team\\
Email: \{\IEEEauthorrefmark{1}jean-francois.couchot,\IEEEauthorrefmark{2}raphael.couturier,\IEEEauthorrefmark{3}michel.salomon\}@univ-fcomte.fr\\
Authors in alphabetical order}
}


\maketitle

\begin{abstract}
Conventional state-of-the-art image steganalysis approaches usually consist of a classifier trained with features provided by rich image models. As both features extraction and classification steps are perfectly embodied in the deep learning architecture called Convolutional Neural Network (CNN), different studies have tried to design a CNN-based steganalyzer. The network designed by \emph{Xu et al.} is the first competitive CNN with the combination Spatial Rich Models (SRM) and Ensemble Classifier (EC) providing detection performances of the same order. In this work we propose  a criterion to choose either the CNN or the SRM+EC method for a given input image. Our approach is studied with three different steganographic spatial domain  algorithms: S-UNIWARD, MiPOD, and HILL, using the Tensorflow computing platform, and exhibits detection capabilities better than each method alone. Furthermore, as SRM+EC and the CNN are both only trained with a single embedding algorithm, namely MiPOD, the proposed method can be seen as an approach for blind steganalysis. In blind detection, error rates are respectively of 16\% for S-UNIWARD, 16\% for MiPOD, and 17\% for HILL  on the BOSSBase with a payload of 0.4~bpp. For 0.1~bpp, the respective corresponding error rates are of 39\%, 38\%, and 41\%, and are always better than the ones provided by SRM+EC.



\end{abstract}

\section{Introduction}

During this past decade many steganographic algorithms have been proposed to hide a secret message inside a cover image. Such embedding schemes can operate in the spatial domain, like for example MiPOD~\cite{7289422}, STABYLO~\cite{DBLP:journals/adt/CouchotCG15}, S-UNIWARD~\cite{HFD14}, HILL \cite{li2014new}, WOW~\cite{conf/wifs/HolubF12}, or HUGO~\cite{conf/ih/PevnyFB10} but also in the frequency domain as J(PEG) counterpart of S-UNIWARD. When designing such an algorithm the objective is to provide an approach that changes the cover image as little as possible. Indeed, the less the cover is modified, the less likely the stego image containing the message is to be detected and thus the more secure the steganographic scheme is. Obviously, assessing the security of steganographic tools has given rise to the dual challenge of detecting hidden information, also called steganalysis. In the case of images, little information is usually available to perform steganalysis, we only assume that the image domain is known.

The wide majority of image steganalysis approaches are two-step. The first stage exhibits useful information on image content by computing a set of features and the second one uses them to train a machine learning tool to distinguish cover images from stego ones. For the first step, different Rich Models (RM) have been proposed for the spatial domain (SRM) \cite{FK2013} and the JPEG one \cite{DBLP:journals/tifs/HolubF15}, while for the second step the most common choice is Ensemble Classifier (EC) \cite{DBLP:journals/tifs/KodovskyFH12}. This combination RM+EC is used in many state-of-the-art image steganalysis tools. As an illustration, in \cite{FK2013} stego images obtained with the steganographic algorithm HUGO have been detected with errors of 13\% and 37\%, respectively, for embedding payloads of 0.4 and 0.1~bpp. These errors were slightly reduced (12\% and 36\%) in \cite{DBLP:journals/tifs/HolubF13} thanks to an improved rich model, and a similar model has been applied in the JPEG domain for stego images obtained with the J-UNIWARD steganographic scheme.

Deep learning \cite{conf/mmsys/Dang-NguyenPCB15,lecun2015deep} has led to breakthrough improvements in various challenging tasks in computer vision, becoming the state-of-the-art for many of them. A key reason for this success
is the current availability of powerful computing platforms, and more particularly GPU-accelerated ones. Among the different network architectures belonging to this family of machine learning methods, Convolutional Neural Networks (CNN) \cite{krizhevsky2012imagenet} are very efficient to solve image classification problems. For example, they achieved the best results on the MNIST problem that consists in the automatic recognition of handwritten digits \cite{wan2013regularization}, or the CIFAR benchmark problems \cite{2015arXiv151107289C}. As steganalysis is a similar problem, since the objective is to classify an input image as either a cover or a stego one, the design of a CNN-based steganalyzer has received increasing attention for the past few years.

From an architecture point of view, a CNN is a feedforward neural network composed of two parts matching exactly the two steps used in conventional steganalysis. More precisely, the first part, called the convolutional part, consists of one or several layers trained to extract feature maps becoming smaller with the layer depth. The second one is composed of some fully-connected layers trained simultaneously to perform the classification task. Hence, a CNN does not only learn how to classify, but also how to automatically find a set of features giving a better representation of the input image thanks to 2D~convolution kernels. A feature map is usually produced by a three-step process: a combination of filtered maps of the previous layer (or the input image for the first layer), a nonlinear processing by a neuron, and finally a size reduction through pooling. Therefore, in the convolutional part the training aims at optimizing the kernel values and the neurons biases. More details can be found for example in \cite{krizhevsky2012imagenet}.

The remainder of this paper proceeds as follows. Section~\ref{sec2} presents related works. We start by describing state-of-the-art steganographic schemes in the spatial domain. This is followed by a survey on previous works done on the use of convolutional networks for image steganalysis. The next section first recalls the CNN architecture designed by Xu {\em et al.} \cite{xu2016structural}. After an experimental study, we focus on why it sometimes fails to detect some stego images. Section~\ref{sec4} is devoted to our main proposal: a criterion to choose the best suited method between the CNN and SRM+EC for a given input image, whose relevance is experimentally assessed. The paper ends with a section that summarizes the contributions and outlines suggestions for future research.


\section{Related Works} \label{sec2} 

\subsection{Steganography}

To be self-sufficient, this article recalls the key ideas of the three most secure steganographic tools, namely S-UNIWARD~\cite{HFD14}, MiPOD~\cite{7289422}, and HILL~\cite{li2014new}.

For each of these algorithms, a distortion function $\rho$ associates to each pixel the cost of modifying it. More formally, for a given cover $X$, let $\rho(X)$ be the matrix whose elements represent the cost of increasing or decreasing by 1 the corresponding pixels. By ranking pixels according to their value in $\rho(X)$, one can compute the set of pixels whose modification induces the smallest detectability. For instance the distortion function $\rho_U$ of S-UNIWARD is defined by:
\begin{equation}
\rho_U(X) = \sum_{k=1}^3\frac{1}{|X \star K^{k}| + \sigma} \star |K^{k}|^{\curvearrowleft},
\label{eq:distortion:uniward}
\end{equation}
where $\star$ is a convolution mirror-padded product, $Y^{\curvearrowleft}$ is the result of a 180~degrees rotation of $Y$, $K^{k}$, $1 \le k \le 3$ are Daubechies-8 wavelet kernels in the three directions, and $\sigma$ is a stabilizing constant. It should be noticed that the multiplicative inverse is element-wise applied. An element of $\rho_U(X)$ is small if and only if there are large variations of large cover wavelet coefficients in the three~directions.

In MiPOD, the distortion function $\rho_M$ is obtained thanks to a probabilistic approach. More precisely, let $\beta$ be the matrix defined as the probabilities to increase by~1 the image pixels. The objective of such a scheme is then to find probabilities which minimize a \emph{deflection coefficient}, $\Sigma \sigma^{-4} \beta^2$, where $\sigma$ is the residual variance matrix of image pixels. Notice that the product is element-wise applied and the sum concerns all the elements of the matrix. Thanks to a Wiener filter and a Lagrangian method, $\beta$ can be computed. Considering such pixel probabilities, the distortion cost $\rho_M$ is defined by: 
\begin{equation}
\rho_M(X) = \ln\left(\frac{1}{\beta}-2\right).
\label{eq:distortion:mipod}
\end{equation}
Again, the multiplicative inverse is applied element-wise.

Finally, the distortion function $\rho_H$ of the HILL steganographic scheme is based on combinations of convolution products. However, contrary to the distortion function $\rho_U$ of S-UNIWARD, this one combines a high-pass filter $H_1$ and two low-pass filters $L_1$ and $L_2$. More precisely, $\rho_H$ is defined by:
\begin{equation}
\rho_H(X) = \frac{1}{|X \star H_1| \star L_1} \star L_2,
\label{eq:distortion:hill}
\end{equation}
where 
$$
H_1 = \left[ \begin{array}{rrr} -1 & 2 & -1 \\ 2 & -4 & 2 \\ -1 & 2 & -1 \end{array} \right]
$$
and $L_1$ (resp. $L_2$) is a $3\times 3$  (resp. $5 \times 5$) mean matrix.   
    
In all aforementioned schemes, the distortion function reflects the underlying image model. Hence, the distortion function $\rho$ returns a large value in a easy-defined or smooth area, whereas in a textured or ``chaotic'' area, \textit{i.e.}, with no model, it returns a small value.



\subsection{CNN-based steganalysis}

The first attempt at designing a CNN-based steganalyzer for image steganalysis is due to Tan {\em et al.} \cite{tan2014stacked}. Their proposal, a stacking of convolutional auto-encoders, yielded for HUGO a detection error more than twice as bad as the one given by SRM+EC: 31\% compared to 14\% for a payload of 0.4~bpp. Despite a low detection efficiency, this work highlighted two interesting points. First, deep learning, and more particularly the convolutional neural network architecture, seems to be a promising concept for image steganalysis. Second, a high-pass filtering of the input image allows to greatly improve the detection performance. This latter point is linked to the kernels used to enhance cover pixel predictors in weighted stego-image analysis \cite{ker2008revisiting} and the linear, and non-linear, high-pass filters producing quantized image noise residuals used to build a rich model of the noise component in~\cite{DBLP:journals/tifs/FridrichK12}.

The following study by Qian {\em et al.} \cite{qian2015deep}, has dealt with $256\times256$ input images, and has proposed a CNN consisting of a convolutional part of 5~layers producing at the end 256~features, which are then processed by a fully-connected part of two hidden layers and a final output one of two~softmax neurons.
The preliminary high-pass filtering is done using a $5\times5$ kernel, called $F_0$, similar to the $5\times5$ kernel predictor obtained in \cite{DBLP:journals/tifs/FridrichK12}. As noticed by Fridrich and Kodovsk\'y in \cite{DBLP:journals/tifs/FridrichK12}, this kernel is inspired by a specific embedding algorithm, namely HUGO, but it worked well for the other steganographic algorithms they tested. The detection performance of this CNN was still slightly lower than the state-of-the-art SRM+EC conventional steganalyzer, but Pibre {\em et al.} \cite{pibre2016deep} improved it thanks to a CNN with a different shape. By reducing the number of layers, but using larger ones resulting thus in more feature maps, they were able to reduce the detection error by more than 16\% for S-UNIWARD at 0.4~bpp. They also emphasized that the high-pass filtering with $F_0$ was mandatory, since they were not able to train a CNN without it. 

In comparison with the work of Pibre {\em et al.}, the CNN we designed in \cite{2016arXiv160507946C} being shallow, was quite different and calling into question some assumptions previously made. On the one hand, we proposed a convolutional part of two layers: a first layer reduced to a single $5\times5$~kernel trained to replace $F_0$, followed by a layer using large kernels (almost as large as the image size). On the other hand, the resulting set of 256~features (for an input image of $512\times512$~pixels) was so discriminating that the fully-connected network doing the classification task could be shortened to the two final softmax~neurons. This CNN is able to detect different steganographic algorithms, working in spatial and frequency domains, almost without any error for a payload of 0.4~bpp. Unfortunately, our work, as well as the one of Pibre {\em et al.}, suffers from a redhibitory drawback: stego images were always obtained by using the same embedding key. In fact, this not recommended ``same embedding key" scenario, since embedding several messages with the same key weakens security, comes from a mistake: the use of C++ embedding simulators from Binghamton DDE Lab website. Consequently, when processing stego images produced with different embedding keys, as expected, the detection performance drops dramatically, far below the one of the state-of-the-art SRM+EC approach. Let us notice that the work by Qian {\em et al.}, described in the previous paragraph, might suffer from the same drawback. 

More recently, the works \cite{xu2016structural} and \cite{xu2016ensemble} by Xu {\em et al.} have shown that CNN-based steganalysis remains competitive with conventional steganalysis. In \cite{xu2016structural} they first proposed a structural design of CNNs for steganalysis that is neither large, nor deep, and learns from noise residuals, since they considered as input image the one issued from high-pass filtering using the kernel $F_0$. The architecture of such convolutional networks, which is the basis of our work presented thereafter, will be described in detail in the next section. The experiments they completed have considered two spatial content-adaptive steganographic algorithms: S-UNIWARD and HILL. They have shown that the performance gained by an ensemble of five CNNs is comparable to the one of SRM+EC. In fact, they trained 5~CNNs because the dataset was split into 5~test subsets, and thus an input image class was predicted by averaging the output probabilities of the CNNs. As an illustration, for S-UNIWARD they observed detection accuracies of 57.33\% and 80.24\% for payloads of 0.1 and 0.4~bpp, whereas the ones of SRM+EC were, respectively, of 59.25\% and 79.53\%. However, the authors have not applied their approach to JPEG domain steganographic algorithms and they pointed out that in the spatial domain more advanced conventional steganalysis methods, such as \cite{tang2016adaptive} or \cite{denemark2016improving}, still outperformed their approach.

In the following work~\cite{xu2016ensemble}, Xu {\em et al.} decided to study the merging of CNNs and ensemble classifier. The background idea is to train a second level classifier using information provided by CNNs. Furthermore, they also sligthly modified the architecture of the original CNN designed in \cite{xu2016structural}, which is denoted by `SIZE~128'. This new `SIZE~256' CNN~architecture, the number of final features given by the convolutional part, has one more layer and changed pooling sizes in the previous ones. In addition to the ensemble method described in the above paragraph, called PROB, where EC will use the output of 16~CNNs instead of five, they defined two further ensemble methods. The first one, called PROB\_POOL, is supposed to lower the loss of information induced by the pooling operation. Indeed, when the stride value is larger than~one, some sampling operations are dropped.  For a stride value $p>1$, applying the pooling on a block of $p \times p$ pixels gives a single value, whereas for a stride of~1 the same block would have been replaced by $p \times p$ values. The idea is thus to also consider independently each remaining $p \times p - 1$ possible sampling, which means as they used pooling operations with stride of~2 that there are 4~possible samplings in a convolutional layer. Consequently, a CNN with a single convolution layer would be applied 4~times on the same input image, giving a prediction for each possible sampling. For two convolutional layers, the number of sampling combinations is $4^2=16$, and so on. In the case of the `SIZE~256' CNN~architecture, since 16~trainings are done for each, the final prediction for an image is obtained by averaging $4^5 \times 16 = 1024 \times 16 = 16384$~predictions. The second new ensemble method, called FEA, is simpler: it uses an architecture merging the convolutional part and the ensemble classifier. 
Let us notice that the larger number of both CNNs and features results in lower bias and variance. From the experiments done with these 6~ensemble scenarios (two size and three methods), Xu {\em et al.} concluded that it might be interesting to replace the fully-connected part of the CNN by EC for image steganalysis. The FEA method was the best one for `SIZE~256' with a detection error of 18.44\%, but, compared to the 18.97\% of PROB without EC, the improvement seems minor.

Another newly published contribution \cite{7532860} by Qian {\em et al.} shows that a CNN trained to detect a spatial steganographic algorithm with a high payload embedding allows to improve the detection performance for lower payloads. For WOW algorithm, the pre-training with stegos obtained using a payload of 0.4~bpp led to lower detection errors in comparison with SRM+EC for 0.1, 0.2, and 0.3~bpp payloads. 

Finally, we can notice the latest work \cite{2016arXiv161103233Z} by Zeng {\em et al.} dealing with JPEG domain steganalysis. A domain which has received considerably less attention when designing CNN-steganalyzers. In this paper, the authors state that compared to rich models, a CNN cannot efficiently learn to extract noise residuals, being unable to find similar or better kernels than the ones used in rich models. 
Therefore they propose to start by manually applying to the input image the first two phases of DCTR \cite{DBLP:journals/tifs/HolubF15}, namely a convolution followed by {\it Q}uantization \& {\it T}runcation ({\it Q\&T}). They use 25~residual images, where each is obtained by using a $5 \times 5$ DCT basis pattern, and three {\it Q\&T} combinations. Then, for each group of residual maps for a given {\it Q\&T} combination, a subnetwork corresponding to a simplified version of the convolutional part proposed by Xu {\em et al.} in \cite{xu2016structural} is trained to produce a feature vector of 512~components. To obtain the final prediction, the three vectors are concatenated and given as input to a three-layer fully connected network, which is trained together with the three subnets. Based on the experiments performed on more or less large databases of images issued from ImageNet, the authors claim that their proposal outperforms all other existing steganalysis approaches (no matter whether they are deep learning or conventional ones).

\section{Convolutional Neural Networks \\
for Image Steganalysis} \label{sec3}

This section begins with the description of the CNN architecture proposed in \cite{xu2016structural}, which is then experimentally studied in order to analyze what causes it to fail or not succeed to detect whether an image embeds a secret message. 

\subsection{The CNN architecture proposed by Xu {\em et al.}}

\begin{figure*}
\centering
\begin{tabular}{cc}
(a) & \parbox{0.98125\linewidth}{\includegraphics[width=0.975\linewidth]{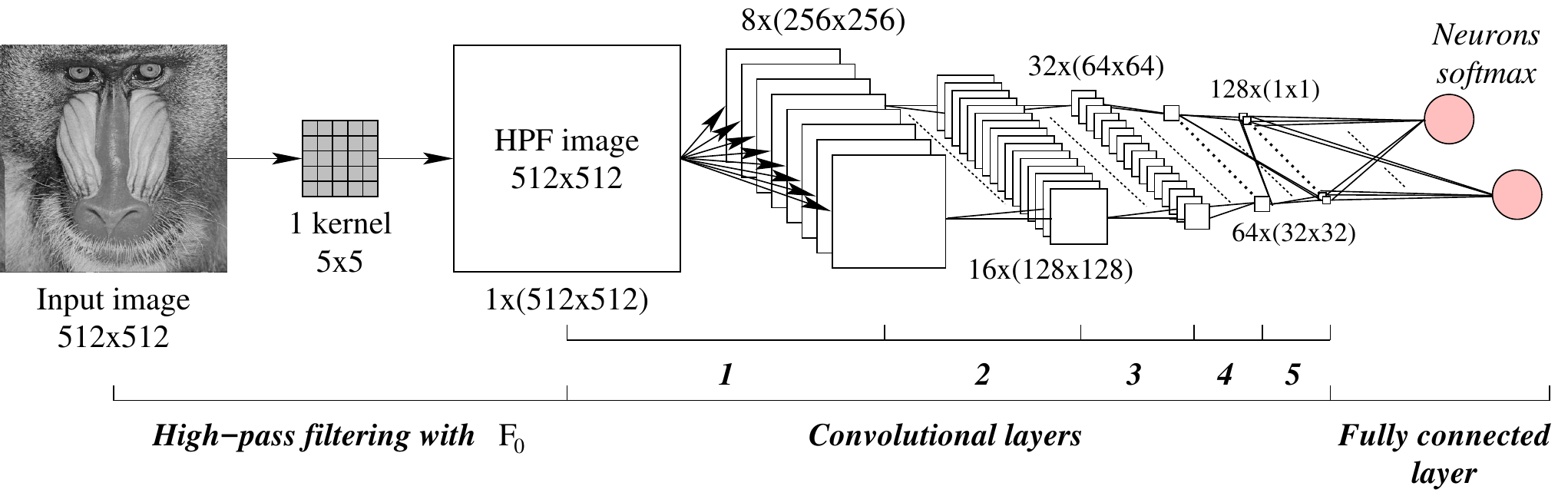}} \\
(b) & \parbox{0.98125\linewidth}{\includegraphics[width=0.975\linewidth]{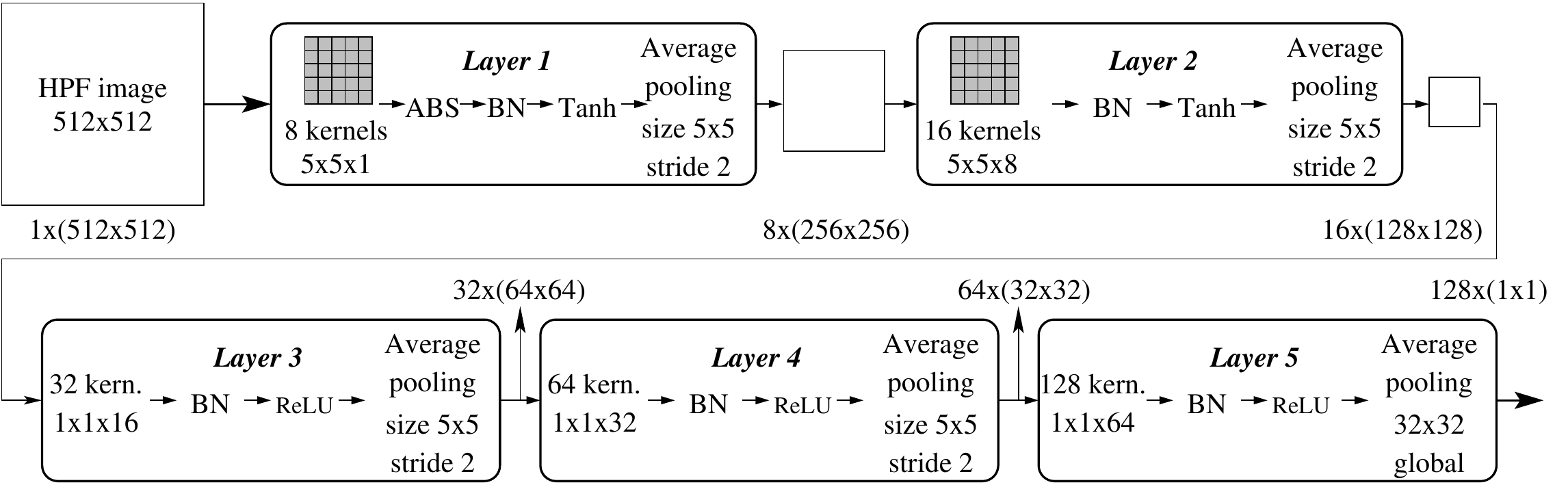}}
\end{tabular}
\caption{CNN proposed by Xu {\em et al.}: (a)~overall architecture and (b)~detailed view. of the convolutional part.}
\label{fig:cnn_xu}
\end{figure*}

Like almost all the previous research works on CNNs for image steganalysis in the spatial domain, and as can be seen in Figure~\ref{fig:cnn_xu}(a), Xu {\em et al.} proposed an architecture that takes as input a high-pass filtered (HPF) version of the input image. Therefore, they used the kernel denoted by $F_0$ in \cite{DBLP:journals/tifs/FridrichK12,qian2015deep,pibre2016deep} in order to highlight noise residuals. This filtering is obviously of great importance, since it provides the input information to the CNN, and thus must be suited to the classification task. The relevance of this kernel comes from its design for rich models. Moreover in their work Pibre {\em et al.} suggest that it is mandatory for CNN-based steganalysis. Overall, the CNN consists of 5~convolutional layers and a fully-connected part reduced to two softmax neurons, each of them giving the probability for an image to belong to one of the two classes (cover or stego). A classification part reduced to output neurons means that a linear classification is able to distinguish covers from stegos using the features produced by the final convolutional layer. We came to a similar classification part in our previous work \cite{2016arXiv160507946C}, even if in our case the problem was simpler due to the use of a single embedding key. In \cite{qian2015deep,pibre2016deep} they had several hidden layers, more or less large (from 128 to 1000~neurons).

Starting with a HPF image of $512 \times 512$~pixels, the convolutional part results in a vector of~128~features, as shown by the detailed view of Figure~\ref{fig:cnn_xu}(b). Each of the four first layers successively halves the image size by generating feature maps using an average pooling, while the fifth one replaces each feature map ($32 \times 32$~pixels) by a single value obtained through a global average pooling. From the convolution kernel point of view, Layers~1 and 2 learn $5 \times 5$ kernels, and the remaining layers $1 \times 1$ ones, the idea being to avoid an overfitting of the CNN to image content and/or stego noise. Layer~1 has also a specific function applied onto the outcome of the convolution, namely the absolute function (ABS), supposed to ensure that the model takes care of the symmetry in noise residuals like in rich models \cite{DBLP:journals/tifs/FridrichK12}. Batch normalization (BN) \cite{ioffe2015batch} is performed in every convolutional layer, because it improves the training (faster and lower prediction error) by allowing a stable distribution of non-linearity inputs. A mixing of {\it Tanh} and {\it ReLU} non-linear activation offered the best performance.

\subsection{Detection performance evaluation of the CNN}

To study and assess the performance of Xu {\em et al.} proposal, which was originally evaluated using a modified version of Caffe toolbox, the corresponding CNN has been implemented with the open source software library TensorFlow \cite{2016arXiv160304467A}. The implementation is available on download from {\tt GitHub}\footnote{\url{https://github.com/rcouturier/steganalysis_with_CNN_and_SRM.git}}. We had previously used in \cite{2016arXiv160507946C} the computing platform Torch \cite{collobert2011torch7}, but TensorFlow seems to make better use of the GPUs. All the experiments are performed on a NVIDIA Tesla~Titan~X GPU, using as cover database the well-known BOSSBase \cite{Bas2011} which consists of 10,000~grayscale images having a size of $512~\times~512$~pixels. Six stego images are associated to each cover image. They are obtained by embedding a message with S-UNIWARD, MiPOD, and HILL algorithms, thanks to the Matlab scripts downloadable from Binghamton DDE Lab website, 
considering two different payload values: 0.1 and 0.4~bpp. 
During a training execution a CNN is trained on a set of 5,000~cover-stego pairs and its detection performance assessed on a the remaining 5,000~pairs. Both training and testing sets are built by randomly picking pairs of images, using a different seed for each execution. 

Notice that even if we implemented exactly a CNN according to the proposal, there is a major difference in comparison with the original work in the way the final prediction is obtained. In \cite{xu2016structural} Xu {\em et al.} generated from a training set five different non-overlapping 4,000/1,000~splits and each of them is used to train separately a CNN occurrence. The purpose of the smaller subset of 1,000~pairs is to serve as a validation set to check when to stop the training process. The final prediction for a given test image is then obtained by averaging the five output probabilities. In \cite{xu2016ensemble}, in addition to a slightly deeper CNN version, the authors investigated different ensemble methods with a larger number of CNNs. 

Let us explain how the final prediction is computed with a set of $T$ trained CNNs which are denoted as $\textrm{CNN}_1, \textrm{CNN}_2, \dots, \textrm{CNN}_T$.
First of all, each $\textrm{CNN}_i$, $1 \le i \le T$, memorizes its $L$ last versions provided by the $L$ last training epochs obtained all along the program execution. These internal CNNs are denoted as $\textrm{CNN}_i^1, \textrm{CNN}_i^2,\dots \textrm{CNN}_i^L$. Each of these internal CNNs gives an answer, which is 0 if the tested image~$I$ is declared as cover and 1 otherwise. Finally, the average of all the values is computed, and a discrete answer is returned by each CNN depending on whether this average is greater or equal to 0.5 or not. This is formalized for each~$i$, $1 \le i \le T$, by:
\begin{equation}
\begin{split}
\textrm{is\_stego}&(I,\textrm{CNN}_i) = \hfill \\
&\left\{
\begin{array}{ll}
0 & \textrm{if }  \left(\frac{1}{L} \sum_{j=1}^{L} \textrm{is\_stego}(I,\textrm{CNN}_i^j) \right) <0.5, \\
\smallskip
1 & \mbox{otherwise.} 
\end{array}
\right.
\end{split}
\end{equation}

The aggregation of these results must take into consideration the fact that an image $I$ we want to classify is used in training step in some $\textrm{CNN}_i$ or not. Let us consider the set 
$$\mathcal{T}_I = \left\{ i | 1 \le i \le T \textrm{ and $I$ is used in testing step of $\textrm{CNN}_i$} \right\}
$$
and $T_I$ be its cardinality. The number $T_I$ counts the number of times $I$ is used as a testing image by some CNNs. The final answer for image $I$ is then the discrete answer of the average of all the CNNs that have used $I$ as testing image. This is formalized by:
\begin{equation}
\begin{split}
\textrm{is\_stego\_CNN}&(I) = \\
&\left\{
\begin{array}{ll}
0 & \textrm{if } \left(\frac{1}{T_I} \sum_{i \in \mathcal{T}_I} \textrm{is\_stego}(I,\textrm{CNN}_i) \right)<0.5, \\
\smallskip
1 & \textrm{otherwise.} 
\end{array}
\right.
\end{split}
\end{equation}
Indeed, as both training and testing sets are built by randomly picking images, the number of times an image $I$ is in a test set varies (being at most equal to $T$).

\begin{figure*}
\centering
\begin{tabular}{cccc}
\includegraphics[scale=0.225]{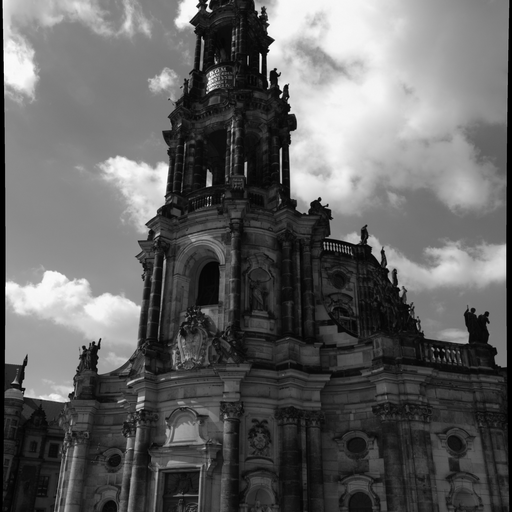} & \includegraphics[scale=0.225]{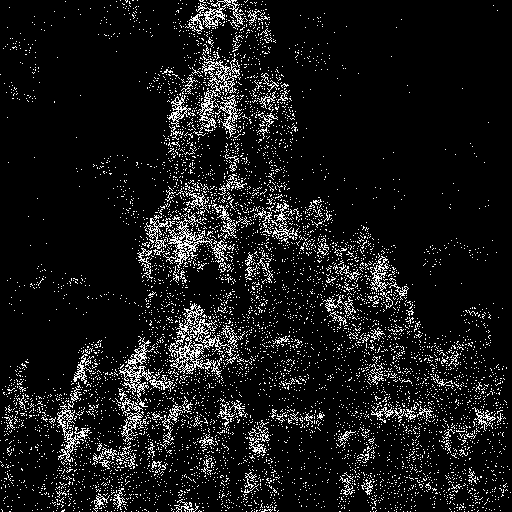} &
  \includegraphics[scale=0.225]{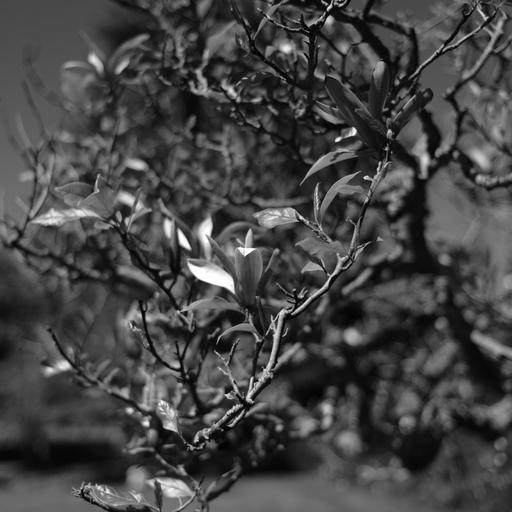} & \includegraphics[scale=0.225]{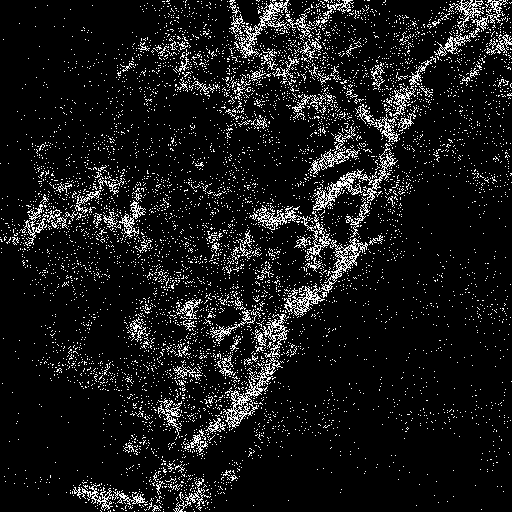} \\
\multicolumn{2}{c}{(a) Cover 1388.pgm and differences with stego.} & \multicolumn{2}{c}{(b) Cover 8873.pgm and differences with stego.} \\
\vspace{0.125cm} \\
\includegraphics[scale=0.225]{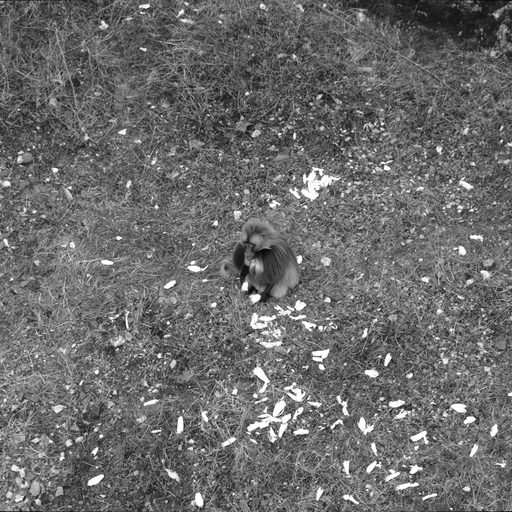} & \includegraphics[scale=0.225]{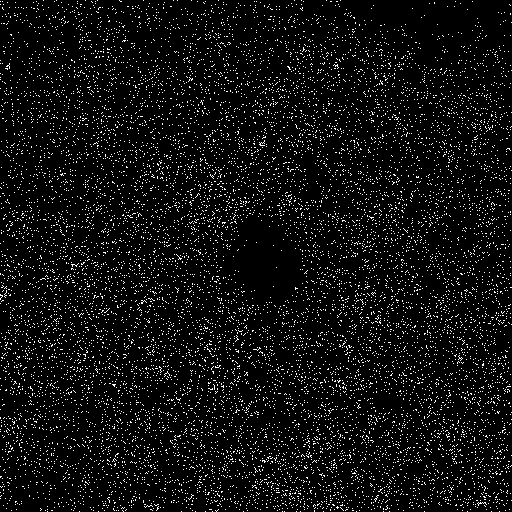} &
  \includegraphics[scale=0.225]{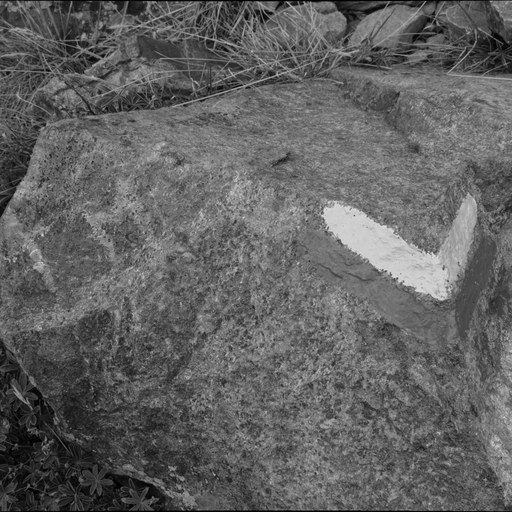} & \includegraphics[scale=0.225]{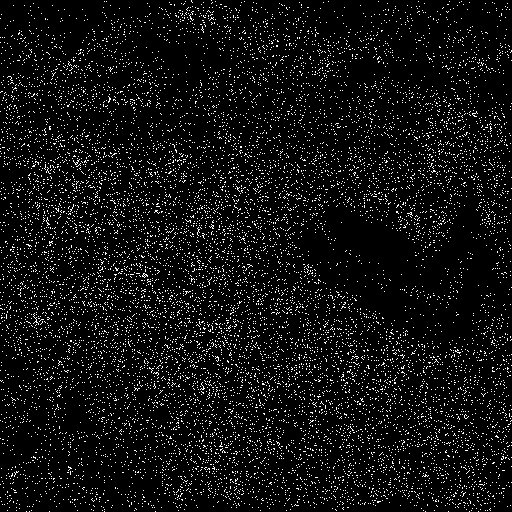} \\
\multicolumn{2}{c}{(c) Cover 1911.pgm and differences with stego.} & \multicolumn{2}{c}{(d) Cover 3394.pgm and differences with stego.}
\end{tabular}
\caption{Examples of differences images between cover and corresponding stego when embedding is performed using MiPOD with a payload of 0.4~bpp.}
\label{fig:examples}
\end{figure*}

Furthermore, due to the huge computation cost we have only trained CNNs using MiPOD dataset and tested them directly on the S-UNIWARD and HILL datasets. Hence, we can also assess whether a CNN trained with an embedding scheme is still competitive for the other ones. We have chosen MiPOD because it is supposed to have the best security and thus should be the most difficult to detect, even if MiPOD and HILL are very close according to~\cite{7289422}.

The key training parameters for reproducible experiments are discussed thereafter. First, a CNN is trained for a maximum number of training epochs $E_{\max}$ set to $1,000$ and $300$, respectively, for embedding payloads of 0.1 and 0.4~bpp, without any overfitting control with a validation set. To compute the prediction given by a network $\textrm{CNN}_i$ for an image~$I$, $L=20$ occurrences are used. These values were chosen after some preliminary runs. Second, the network parameters are optimized by applying a mini-batch stochastic gradient descent, a typical choice in deep learning. Batch normalization is in particular known to make the parameters updates more stable. In our experiments, we used a mini-batch size of $64$~samples, but without ensuring that both cover and stego of a given pair are in the same batch like Xu {\em et al.}. The gradient descent also uses parameters values which are almost similar to the original ones: a learning rate initialized to $0.001$, but with no weight decay, and a momentum set to $0.9$.

The obtained average detection errors are reported in Table~\ref{tab:init_errors}. The first line labelled with "Caffe~\cite{xu2016structural}" recalls values given in~\cite{xu2016structural}. 
The second line gives the average error rates from $T=12$~independent training runs of the TensorFlow implementation for embedding payload of $0.4$~bits per pixel. The third gives the average error rates for 200~runs with classical SRM+EC. In this latter context maxSRMd2~\cite{DBLP:conf/wifs/DenemarkSHCF14} has been used as a feature set. Finally, the last line gives the results obtained when the training stage is executed with images modified by MiPOD, whereas the testing stage is executed with images modified with another embedding scheme. The classifier is SRM+EC and the feature set is maxSRMd2. Remember that an objective of this study is to develop a completely blind steganalysis approach. It is not difficult to understand that the detection error is larger in this context, since the steganographic scheme used to learn is not necessarily the one used in the testing phase.

\begin{table}[ht]
\centering
\caption{Average detection error as a function of classifier (original Caffe by Xu {\em et al.}, our TensorFlow trained only with MiPOD at~0.1~/~0.4~bpp, and SRM+EC) and of payload for different steganographic~algorithms.}
\begin{tabular}{|c|p{0.6cm}p{0.6cm}|p{0.6cm}p{0.6cm}|p{0.6cm}p{0.6cm}|}
\cline{2-7}
\multicolumn{1}{c|}{} & \multicolumn{2}{c|}{S-UNIWARD} & \multicolumn{2}{c|}{MiPOD} & \multicolumn{2}{c|}{HILL} \\
\multicolumn{1}{c|}{} & 0.1 & 0.4 &  0.1 & 0.4  & 0.1 & 0.4\\
\hline
Caffe~\cite{xu2016structural}& 42.67 & 19.76 & X  & X & 41.56 & 20.76 \\
\hline
TensorFlow & X & 20.52 & X &  19.36 & X & 20.25  \\
\hline 
SRM + EC & 39.84 & 18.06 & 41.18  & 21.42 & 42.96 & 23.31\\
\hline  
SRM + EC  (blind)& 40.57 & 20.85 & 41.18  & 21.42 & 43.35 & 23.99\\
\hline  
\end{tabular}
\label{tab:init_errors}
\end{table}

From the values given in this table we can draw several conclusions. First, despite the differences highlighted previously, the TensorFlow implementation produces nearly the same performance for S-UNIWARD and HILL than the original Caffe one. Second, we observe that the steganalysis scheme with maxSRMd2 features results in the best performances for S-UNIWARD in case of non-blind steganalysis. Third, for MiPOD the CNN approach is still competitive with SRM+EC. Fourth, the CNNs trained by only making use of the MiPOD dataset can provide a similar detection accuracy for S-UNIWARD and HILL. Obviously, the lowest detection error is gained for the embedding scheme which has provided the training data. Fifth, CNNs outperform SRM+EC in blind steganalysis context, which means that CNNs allow a better generalization to different steganographic algorithms. We are then left to study for which images the CNN proposal by Xu {\em et al.} fails.






\subsection{Characterizing the mis-CNN-classified images}

Let us start with some illustrative examples of images describing the typical behavior of the CNN in the case of MiPOD with payload~0.4~bpp. Figure~\ref{fig:examples} presents four case examples where for each we have the cover image and the corresponding differences between it and the stego one. As can be seen from the images showing the differences, we can distinguish two groups of images according to the pixels modified by the embedding process. On the one hand, it clearly appears that for both images shown on the upper line, 1388.pgm and 8873.pgm, MiPOD modifies mainly pixels corresponding to edges. On the other hand, for 1911.pgm and 3394.pgm, the changes resulting from embedding are scattered without obviously highlighting any underlying image edge. Consequently, since a CNN mainly learns to detect underlying edges, one can easily guess that the CNN-steganalyzer is able to detect both cover and stego for 1388.pgm and 8873.pgm, whereas it fails for the two other images.

We are then left to provide a metric on images which reflects the difficulty to perform the CNN classification task. This metric should in particular allow to cluster the images according to the observations made on Figure~\ref{fig:examples}. 

\begin{figure}[t]
\centering
\includegraphics[width=\columnwidth]{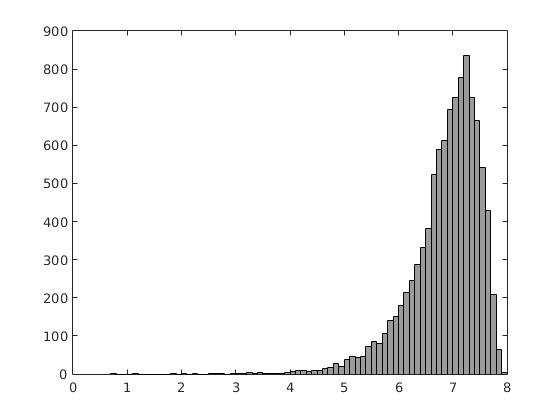}
\caption{Distribution of the BossBase images with respect to entropy.}
\label{fig:entropy_hist}
\end{figure}

As previously mentioned, the distortion value of a pixel is large if this one belongs to an easy-modeled area, \textit{i.e.}, when very little data are required to reproduce this area. The Shanonn entropy $H(I) = -\sum_i p_i \log_2 (p_i)$ is one of the functions that returns the number of bits which are necessary to recompute the image, where $p_i$ is the normalized histogram of pixel values of image $I$. On the BossBase, the entropy ranges in $[0.7, 8]$ and its distribution is represented in Figure~\ref{fig:entropy_hist}.

\begin{figure}[t]
  \begin{center}
    \subfigure[Detection error w.r.t image entropy for SRM+EC.]{
      \begin{minipage}{0.96\columnwidth}
        \begin{center}
            \includegraphics[width=0.99\textwidth]{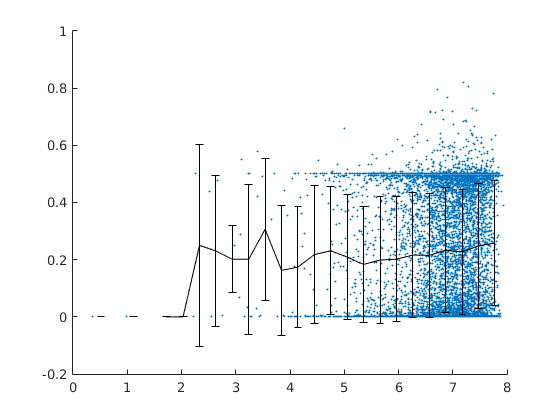}
            \end{center}
      \end{minipage}
      \label{fig:srm_result_wrt_entropy}
    }
           
    \subfigure[Detection error w.r.t image entropy for the CNN by {\em Xu et al.}]{
      \begin{minipage}{0.96\columnwidth}
        \begin{center}
            \includegraphics[width=0.99\textwidth]{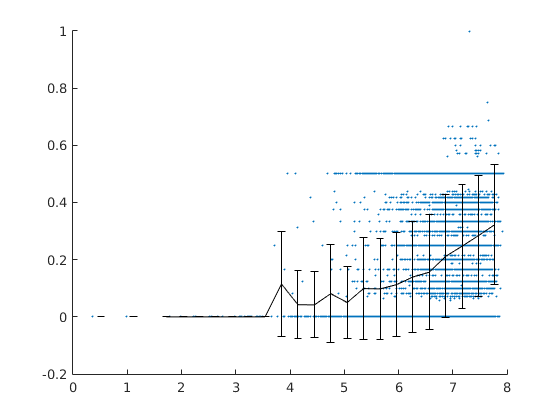}
            \end{center}
      \end{minipage}
      \label{fig:CNN_result_wrt_entropy}
    }
  \end{center}
  \caption{Relation between testing errors and image entropy for embedding with MiPOD 0.4~bpp.}
  \label{fig:entropy_and_error}
\end{figure}

We must now study whether the image entropy is a relevant metric. Therefore, for each image~$I$ of the BossBase we have computed its entropy $H(I)$ and performed 200~classification procedures with SRM+EC (thanks to maxSRMd2) for the embedding algorithm MiPOD at payload 0.4~bpp. The probability $\overline{e_{\textit{SRM+EC}}}(I)$ represents the average testing error for image $I$ when it is used in the testing set. We have also partitioned the entropy interval $[0.7,8]$ into 25~equidistant classes. Figure~\ref{fig:srm_result_wrt_entropy} presents the resulting scatter plot of $(H(I),\overline{e_{\textit{SRM+EC}}}(I))$ pairs and the curve linking the mean error of each class, whereas the bar displays its corresponding standard deviation. Similarly, Figure~\ref{fig:CNN_result_wrt_entropy} shows the scatter plot, curve, and error bars, for the CNN. In that case, $\overline{e_{\textit{CNN}}}(I)$ is the average testing error obtained after training 12~independent networks. This low number explains why, in comparison with the SRM+EC steganalysis context, the points are less vertically spread. A first conclusion of these experiments is that the detection error of SRM+EC seems to be quite independent of the entropy, whereas the one issued from CNN seems to increase as the entropy does. However most of the images have an entropy in $[6,8]$ (as shown in Figure~\ref{fig:entropy_hist}) and it is hard to extract from Figures~\ref{fig:srm_result_wrt_entropy} and \ref{fig:CNN_result_wrt_entropy} a specific behavior inside this interval.



\begin{figure*}[p!]
  \begin{center}
    \subfigure[Detection error w.r.t image $\overline{\rho_U}$ value for SRM+EC.]{
      \begin{minipage}{0.97\columnwidth}
        \begin{center}
            \includegraphics[width=0.99\textwidth]{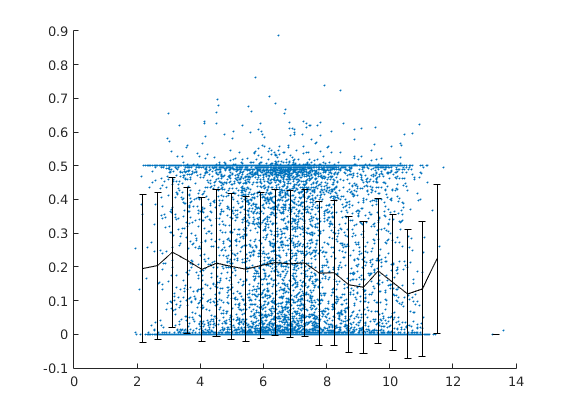}
            \end{center}
      \end{minipage}
      \label{fig:srm_result_wrt_rho_u}
    }
    \hfill
    \subfigure[Detection error w.r.t image $\overline{\rho_U}$ value for the CNN by {\em Xu et al.}]{
      \begin{minipage}{0.97\columnwidth}
        \begin{center}
            \includegraphics[width=0.99\textwidth]{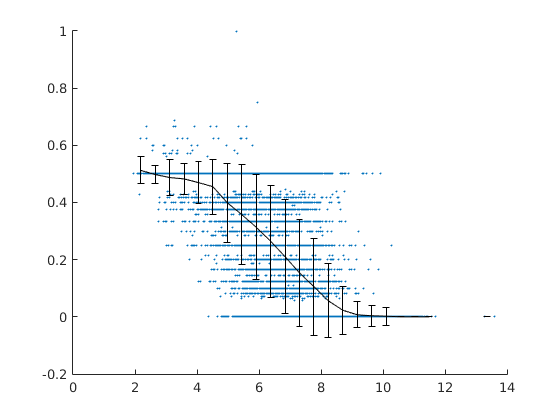}
            \end{center}
      \end{minipage}
      \label{fig:CNN_result_wrt_rho_u}
    }
  
      \subfigure[Detection error w.r.t image $\overline{\rho_H}$ value for SRM+EC.]{
      \begin{minipage}{0.97\columnwidth}
        \begin{center}
            \includegraphics[width=0.99\textwidth]{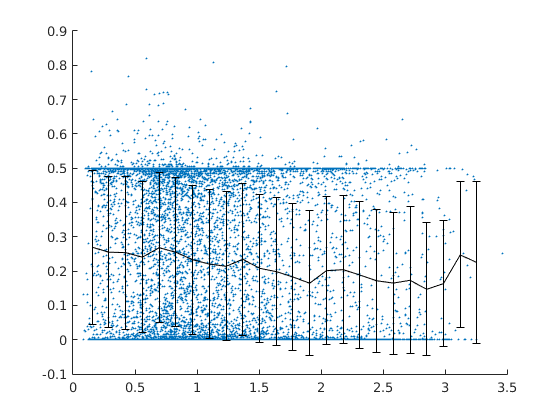}
            \end{center}
      \end{minipage}
      \label{fig:srm_result_wrt_rho_h}
    }
    \hfill
    \subfigure[Detection error w.r.t image $\overline{\rho_H}$ value for the CNN by {\em Xu et al.}]{
      \begin{minipage}{0.97\columnwidth}
        \begin{center}
            \includegraphics[width=0.99\textwidth]{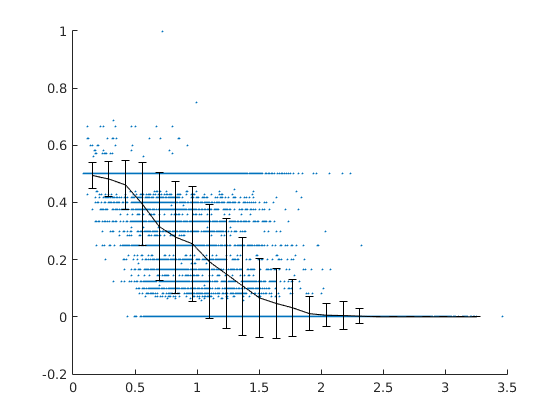}
            \end{center}
      \end{minipage}
      \label{fig:CNN_result_wrt_rho_h}
    }
    
    \subfigure[Detection error w.r.t image $\overline{\rho_M}$ value for SRM+EC.]{
      \begin{minipage}{0.97\columnwidth}
        \begin{center}
            \includegraphics[width=0.99\textwidth]{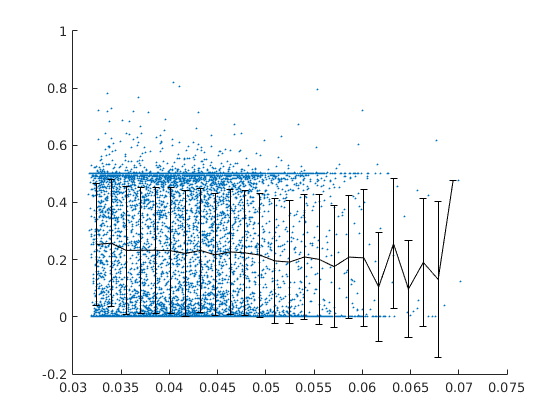}
            \end{center}
      \end{minipage}
      \label{fig:srm_result_wrt_rho_m}
    }
        \hfill
    \subfigure[Detection error w.r.t image $\overline{\rho_M}$ value for the CNN by {\em Xu et al.}]{
      \begin{minipage}{0.97\columnwidth}
        \begin{center}
            \includegraphics[width=0.99\textwidth]{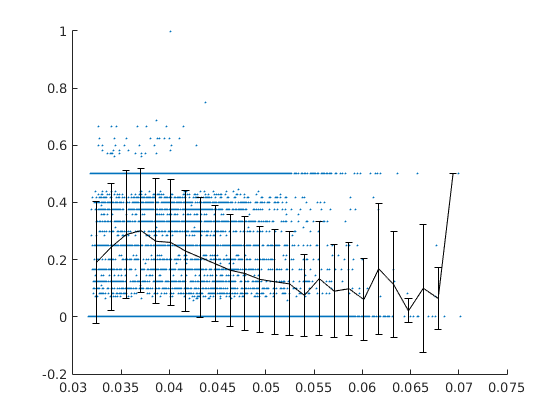}
            \end{center}
      \end{minipage}
      \label{fig:CNN_result_wrt_rho_m}
    }
  \end{center}
  \caption{Relation between testing errors and distortion function mean.}
  \label{fig:rho_and_error}
\end{figure*}

Since the aforementioned steganographic schemes have their own distortion function $\rho$, we have then studied whether another metric can be deduced from it. 
In the same context as the one above, on the one hand,  
Figures~\ref{fig:srm_result_wrt_rho_u}, Figures~\ref{fig:srm_result_wrt_rho_h}, and
Figures~\ref{fig:srm_result_wrt_rho_m}
display the scatter plots, the curve, and error bars for 
$(\overline{\rho_U}(I),\overline{e_{\textit{SRM+EC}}}(I))$,
$(\overline{\rho_H}(I),\overline{e_{\textit{SRM+EC}}}(I))$,
$(\overline{\rho_M}(I),\overline{e_{\textit{SRM+EC}}}(I))$ respectively.
On the other hand, Figures~\ref{fig:CNN_result_wrt_rho_u}, Figures~\ref{fig:CNN_result_wrt_rho_h}, and
Figures~\ref{fig:CNN_result_wrt_rho_m}
display the scatter plots, the curve, and error bars for 
$(\overline{\rho_U}(I),\overline{e_{\textit{CNN}}}(I))$,
$(\overline{\rho_H}(I),\overline{e_{\textit{CNN}}}(I))$,
$(\overline{\rho_M}(I),\overline{e_{\textit{CNN}}}(I))$ respectively.

The scalar $\overline{\rho_U}(I)$ is the mean of all the matrices $\rho_U(X)$ presented in equation~(\ref{eq:distortion:uniward}), where $U$ means S-UNIWARD. The scalar $\overline{\rho_M}(I)$ has a similar definition for MiPOD. Finally $\overline{\rho_H}(I)$ is not directly the mean of all the matrices $\rho_H(X)$ of HILL. Due to its definition (Eq.~(\ref{eq:distortion:hill})), some extremely large values may indeed result from an extremely small denominator. This would lead to a meaningless mean value. To avoid this behavior, extremely large values are excluded from the mean computation.

By focusing on Figures~\ref{fig:srm_result_wrt_rho_u},~\ref{fig:srm_result_wrt_rho_h}, and~\ref{fig:srm_result_wrt_rho_m}, it can be first deduced that the detection error of SRM+EC is quite independent of the value of $\overline{\rho}$. Secondly, considering Figures~\ref{fig:CNN_result_wrt_rho_u} and~\ref{fig:CNN_result_wrt_rho_h}, we can deduce that the CNN testing error continuously decreases with respect to $\overline{\rho_U}(I)$ and with $\overline{\rho_H}(I)$. This behavior is not observed in Figure~\ref{fig:CNN_result_wrt_rho_m}. The good correlation between the prediction accuracy of the CNN for a given image $I$ and the value of $\overline{\rho}(I)$ can be observed in the two former cases but not in the last one. The functions $\overline{\rho_U}$ and $\overline{\rho_H}$ are thus an indicator of the CNN accuracy.
For instance, in Figure~\ref{fig:examples}, for the misclassified images we obtain $\overline{\rho_U}(1911)=2.1$ and $\overline{\rho_U}(3394)=3.06$; on the other hand for the well detected images we get $\overline{\rho_U}(1388)=7.05$ and $\overline{\rho_U}(8874)=7.39$. Thus $\overline{\rho_U}$ and $\overline{\rho_H}$ enable to cluster the images in two groups which are in accordance with those noticed at the beginning of the section.

\section{Taking the best from CNN and SRM+EC predictions to improve detection performance} \label{sec4}

\subsection{Choosing the best method for a given input image}

We have shown in the previous section that the lower the distortion function mean $\overline{\rho_U}$ of an input image is, the more difficult it will be for the CNN to correctly detect whether the image is a cover or a stego. Conversely, SRM+EC gives rather regular detection errors, without showing too much sensitivity to $\overline{\rho_U}$, being robust against the image structure. A look at Figures~\ref{fig:srm_result_wrt_rho_u} and \ref{fig:CNN_result_wrt_rho_u} shows that we can take advantage from these different behaviors to improve the detection performance on the BossBase.

\begin{figure}[t]
\centering
\includegraphics[width=0.95\columnwidth]{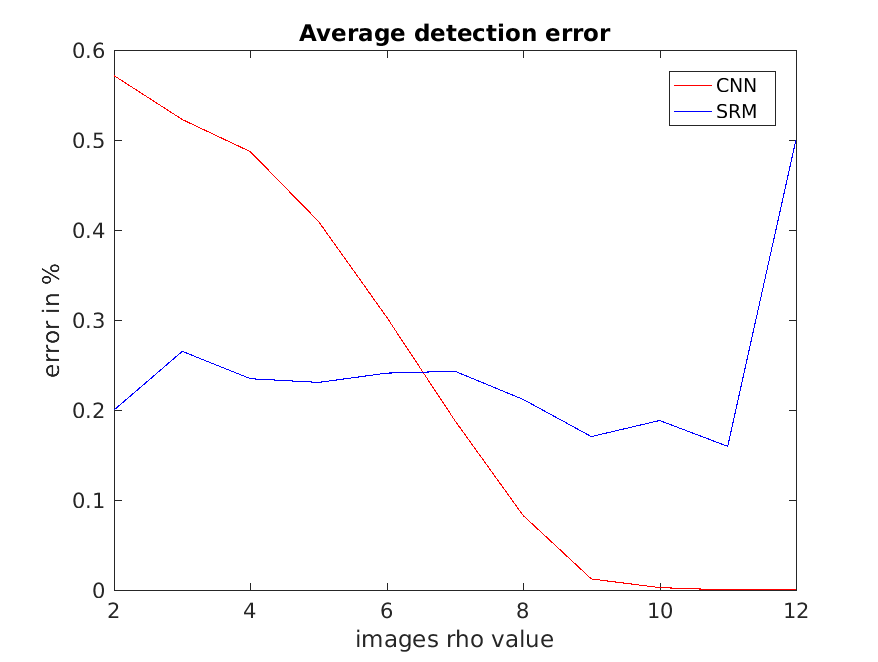}
\caption{Average error of CNN and SRM+EC for MiPOD~0.4~bpp w.r.t $\overline{\rho_U}$.}
\label{fig:cnn_srm_error_mipod0.4}
\end{figure}

In fact, SRM+EC and the CNN can be combined due to complementary purposes. This appears clearly by superimposing the curves of both figures. As can be seen in Figure~\ref{fig:cnn_srm_error_mipod0.4}, from the largest $\overline{\rho_U}$ value up to the point where both curves intersect the CNN is the most competitive, whereas after, towards the lowest $\overline{\rho_U}$ value, it is SRM+EC which is the most accurate. Formally, this can be expressed as follows for an input image~$I$, once $\overline{\rho_U}(I)$ is computed:
\begin{equation}
\left\{ 
\begin{array}{l}
  \mbox{if } \overline{\rho_U}(I)<\overline{\rho_U^{\cap}} \mbox{ use SRM+EC prediction}, \\
  \mbox{otherwise use CNN prediction}
\end{array}
\right.
\end{equation}
where $\overline{\rho_U^{\cap}}$ is the value corresponding to the intersection. For Figure~\ref{fig:cnn_srm_error_mipod0.4}, which deals with the embedding algorithm MiPOD using a payload of 0.4~bpp, we have obtained for the intersection $\overline{\rho_U^{\cap}}=6.6$. Let us emphasize that the same approach can be applied to both S-UNIWARD and HILL steganographic algorithms, leading to different values for $\overline{\rho_U^{\cap}}$.

\begin{figure}[t]
\centering
\includegraphics[width=0.95\columnwidth]{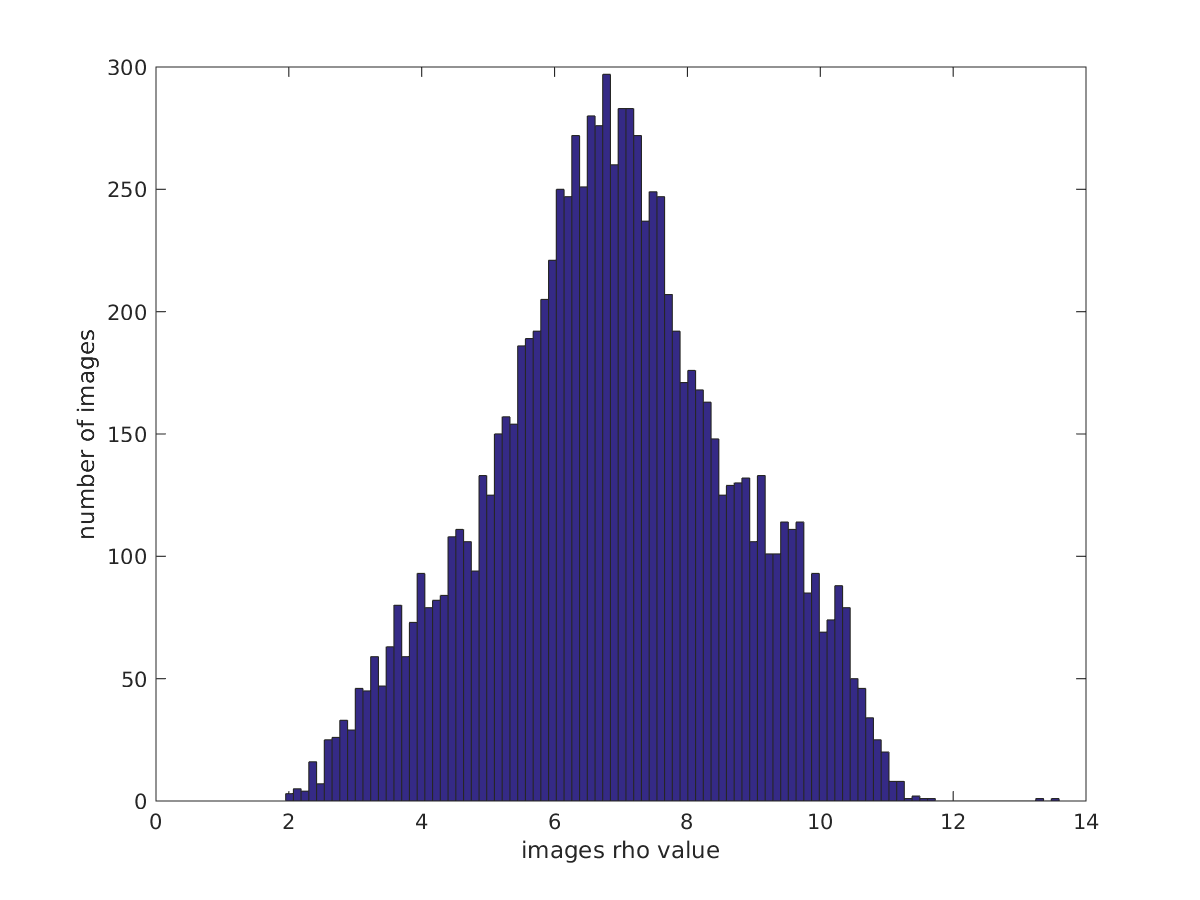}
\caption{Distribution of the BossBase images with respect to $\overline{\rho_U}$}
\label{fig:rho_hist}
\end{figure}

Overall, the feature set generated by a spatial rich model is so large and diverse that it is able to give predictions yielding almost the same level of accuracy, regardless of the pixels modified by the embedding process. Moreover, the computing of the features is precisely defined. Conversely, the CNN learns to extract a set of features to fulfill its classification task according to the data given during the training step. Therefore, it will be well-suited to process images having the same kind of embedding than the main trend in the training set. In other words, images having low $\overline{\rho_U}$ values are so underrepresented in the BossBase that they have a limited influence during the training process. This raises the question of whether the BossBase might be extended to include a larger number of similar images and thus flatten the distribution shown in Figure~\ref{fig:rho_hist}. However, one might wonder if this would effectively improve the CNN detection performance or rather lead to a smoothing of the errors. Furthermore, we have seen that the embedding can be done in quite a different way according to $\overline{\rho_U}$ that the training of a single CNN giving relevant predictions might be far more difficult.

\subsection{Detection performance evaluation of the proposal}

Tables~\ref{tab:new_errors:0.4} and \ref{tab:new_errors:0.1} present in their last column the average detection error obtained using our approach for the three steganographic algorithms and payloads of 0.4~bpp and 0.1~bpp, respectively. In each table the first column gives the performance of SRM+EC computed on images $I$ such that $\overline{\rho_U}<\overline{\rho_U^{\cap}}$, this last value is shown in the second column, while the third column shows the results gained from CNN for the remaining images. We can observe that for each embedding algorithm the proposal improves the detection performance. For an embedding payload of 0.4~bpp, S-UNIWARD has the lowest error rate with 14.82\%, whereas for MiPOD and HILL we have values slightly below 17\%. The lines labelled as non blind correspond to situations where SRM+EC was trained with the same algorithm than the one used to perform the embedding process. Conversely, the lines denoted as blind mean that SRM+EC was trained with MiPOD and then used to detect S-UNIWARD or HILL. This also explains why for both blind and non blind situations the CNN gives the same error when both cases use the same value for $\overline{\rho_U^{\cap}}$. For the lower payload of 0.1~bpp, using $T=9$ independent runs, the improvements provided by our method are also clearly visible, whatever the context of the detection, whether it is blind or~not. 

These results are also somewhat surprising, since one should remember that they are obtained by training only CNNs using images embedding hidden messages with MiPOD. Furthermore, this means that even if each steganographic algorithm has its own distortion function to define which pixels it will modify to embed a message, there is certainly a high redundancy among the modifications made by S-UNIWARD, HILL, and MiPOD on the same cover image. 



\begin{table}[ht]
\centering
\caption{Average detection error according to $\overline{\rho_U^{\cap}}$ for different steganographic algorithms with embedding payload of 0.4~bpp.}
\begin{tabular}{|c|c|c|c|c|}
\hline
 & SRM+EC & $\overline{\rho_U^{\cap}}$ & CNN & CNN~+~SRM+EC \\
\hline
S-UNIWARD & \multirow{2}{*}{20.01} & \multirow{2}{*}{7.1} & \multirow{2}{*}{8.25} & \multirow{2}{*}{14.82} \\
non blind & & & & \\
\hline
S-UNIWARD & \multirow{2}{*}{22.05} & \multirow{2}{*}{6.9} & \multirow{2}{*}{9.5} &  \multirow{2}{*}{15.87} \\
 blind & & & & \\
\hline
MiPOD & \multirow{2}{*}{23.89} & \multirow{2}{*}{6.6} & \multirow{2}{*}{9.26} & \multirow{2}{*}{15.65} \\
non blind & & & & \\
\hline
HILL & \multirow{2}{*}{24.51} & \multirow{2}{*}{6.6} & \multirow{2}{*}{9.78} & \multirow{2}{*}{16.22} \\
non blind & & & & \\
\hline
HILL & \multirow{2}{*}{25.41} & \multirow{2}{*}{6.6} & \multirow{2}{*}{9.78} & \multirow{2}{*}{16.61} \\
 blind &  &     &       &       \\
\hline
\end{tabular}
\label{tab:new_errors:0.4}
\end{table}

\begin{table}[ht]
\centering
\caption{Average detection error according to $\overline{\rho_U^{\cap}}$ for different steganographic algorithms with embedding payload of 0.1~bpp.}
\begin{tabular}{|c|c|c|c|c|}
\hline
 & SRM+EC & $\overline{\rho_U^{\cap}}$ & CNN & CNN~+~SRM+EC \\
\hline
S-UNIWARD & \multirow{2}{*}{40.08} & \multirow{2}{*}{9.2} & \multirow{2}{*}{23.36} & \multirow{2}{*}{38.06} \\
non blind & & & & \\
\hline
S-UNIWARD & \multirow{2}{*}{41} & \multirow{2}{*}{9.2} & \multirow{2}{*}{23.36} & 
\multirow{2}{*}{38.88} \\
 blind & & & & \\
\hline
MiPOD & \multirow{2}{*}{42.13} & \multirow{2}{*}{8} & \multirow{2}{*}{25.84} & \multirow{2}{*}{37.82} \\
non blind & & & & \\
\hline
HILL & \multirow{2}{*}{43.48} & \multirow{2}{*}{8.9} & \multirow{2}{*}{21.88} & \multirow{2}{*}{40.24} \\
non blind & & & & \\
\hline
HILL & \multirow{2}{*}{44.30} & \multirow{2}{*}{8.3} & \multirow{2}{*}{27.72} & \multirow{2}{*}{40.64} \\
blind & & & & \\
\hline
\end{tabular}
\label{tab:new_errors:0.1}
\end{table}

A closer look on the performances of each steganalyzer on the subset of images it has to classify according to $\overline{\rho_U^{\cap}}$ explains why our proposal is relevant. Indeed, in comparison with the performances shown in Table~\ref{tab:init_errors} we can remark that the SRM+EC error rate is slightly worse than on the whole dataset. Thus, we take advantage from the low error rate of the CNN at a price of a slightly worse misclassification by SRM+EC. Another point to notice is the evolution in opposite directions of $\overline{\rho_U^{\cap}}$ and payload values, which means that, as expected, the scatterness of the modified pixels increases and thus is more difficult to detect with the current CNN architecture. Nevertheless, our approach allows us to build a competitive blind steganalyzer, which gives lower detection errors than CNN based only or SRM+EC based only approaches.

\section{Conclusion and future work}

Over the past two years the design of deep learning based approaches for image steganalysis in spatial domain, using more particularly convolutional neural networks, has received an increasing attention due to their impressive successes on many classification tasks. Earlier works were not able to reach the detection performance of the conventional steganalysis approaches using rich models. Nonetheless they showed that such deep neural networks are relevant for the design of steganalyzers. Recently, Xu {\em et al.} have introduced a CNN architecture, which, to the best of our knowledge, is the most competitive one compared to rich models with ensemble classifier. In this paper, rather than designing a further new CNN architecture, we have investigated when this CNN architecture fails in order to propose a method allowing to improve the detection performance on the BossBase for different spatial steganography algorithms.

Thanks to a TensorFlow implementation of the CNN, giving nearly the same detection performance than the original Caffe one for S-UNIWARD and HILL, we have found a metric strongly correlated with the CNN classification performance. This metric consists in the mean of all the elements in the cost matrix provided by the distortion function~$\rho$ of the considered steganographic algorithm for a given input image. We have shown that the lower this latter value $\overline{\rho_U}$ for S-UNIWARD is, the more the CNN fails to correctly detect if the image is a cover or a stego. Fortunately, the CNN and SRM+EC detection errors evolve in different ways according to the metric function, where rich models offer lower error rates when it decreases. By computing the intersection of the corresponding curves we are then able to define a reliable criterion allowing to decide, for an input image, when to use the CNN or SRM+EC to obtain the most accurate prediction. The experiments done considering the steganographic algorithms S-UNIWARD, HILL, and MiPOD, have validated the proposed criterion, since it has always led to improved detection performances, regardless of the embedding payload value. Let us also emphasize another contribution of this work which is to have designed a steganalyzer insensitive to the embedding process (blind detection). Even if the considered CNN is trained with a specific steganographic algorithm, MiPOD in our case, other embedding methods can be detected with a similar accuracy. We also observed that the SRM+EC is quite efficient in blind situations. 

Our future work will focus on two aspects. First, it might be interesting to subdivide the BossBase in disjoint subsets according to the average distortion function value and to train several CNNs on them. However, to be able to train a CNN for low $\overline{\rho_U}$ values, the database should be expanded in order to include more images corresponding to this case of study. Second, CNNs dealing with spatial domain steganalysis work on a single high-pass filtered version of the input image, but since this filtering mainly highlights edges it does not give a relevant information to the CNN when the embedding is not edge-based. Therefore, we plan to replace the single filter by a filter bank, an approach which in the case of the JPEG domain steganalysis seems to be successful according to \cite{2016arXiv161103233Z}.




\section{Acknowledgments}

This article is partially funded by the Labex ACTION program (ANR-11-LABX-01-01 contract) and the Franche-Comt\'e regional council. We would like to thank NVIDIA for hardware donation under CUDA Research Center 2014 and the M\'esocentre de calcul de Franche-Comt\'e for the use of the GPUs.

\bibliographystyle{IEEEtran}
\bibliography{references}

\begin{thebibliography}{10}
\providecommand{\url}[1]{#1}
\csname url@samestyle\endcsname
\providecommand{\newblock}{\relax}
\providecommand{\bibinfo}[2]{#2}
\providecommand{\BIBentrySTDinterwordspacing}{\spaceskip=0pt\relax}
\providecommand{\BIBentryALTinterwordstretchfactor}{4}
\providecommand{\BIBentryALTinterwordspacing}{\spaceskip=\fontdimen2\font plus
\BIBentryALTinterwordstretchfactor\fontdimen3\font minus
  \fontdimen4\font\relax}
\providecommand{\BIBforeignlanguage}[2]{{%
\expandafter\ifx\csname l@#1\endcsname\relax
\typeout{** WARNING: IEEEtran.bst: No hyphenation pattern has been}%
\typeout{** loaded for the language `#1'. Using the pattern for}%
\typeout{** the default language instead.}%
\else
\language=\csname l@#1\endcsname
\fi
#2}}
\providecommand{\BIBdecl}{\relax}
\BIBdecl

\bibitem{7289422}
V.~Sedighi, R.~Cogranne, and J.~Fridrich, ``Content-adaptive steganography by
  minimizing statistical detectability,'' \emph{IEEE Transactions on
  Information Forensics and Security}, vol.~11, no.~2, pp. 221--234, Feb 2016.

\bibitem{DBLP:journals/adt/CouchotCG15}
\BIBentryALTinterwordspacing
J.~Couchot, R.~Couturier, and C.~Guyeux, ``{STABYLO:} steganography with
  adaptive, bbs, and binary embedding at low cost,'' \emph{Annales des
  T{\'{e}}l{\'{e}}communications}, vol.~70, no. 9-10, pp. 441--449, 2015.
  [Online]. Available: \url{http://dx.doi.org/10.1007/s12243-015-0466-7}
\BIBentrySTDinterwordspacing

\bibitem{HFD14}
\BIBentryALTinterwordspacing
V.~Holub, J.~Fridrich, and T.~Denemark,
  ``\BIBforeignlanguage{English}{Universal distortion function for
  steganography in an arbitrary domain},''
  \emph{\BIBforeignlanguage{English}{EURASIP Journal on Information Security}},
  vol. 2014, no.~1, 2014. [Online]. Available:
  \url{http://dx.doi.org/10.1186/1687-417X-2014-1}
\BIBentrySTDinterwordspacing

\bibitem{li2014new}
B.~Li, M.~Wang, J.~Huang, and X.~Li, ``A new cost function for spatial image
  steganography,'' in \emph{2014 IEEE International Conference on Image
  Processing (ICIP)}.\hskip 1em plus 0.5em minus 0.4em\relax IEEE, 2014, pp.
  4206--4210.

\bibitem{conf/wifs/HolubF12}
V.~Holub and J.~J. Fridrich, ``Designing steganographic distortion using
  directional filters.'' in \emph{WIFS}.\hskip 1em plus 0.5em minus 0.4em\relax
  IEEE, 2012, pp. 234--239.

\bibitem{conf/ih/PevnyFB10}
\BIBentryALTinterwordspacing
T.~Pevn{\'y}, T.~Filler, and P.~Bas, ``Using high-dimensional image models to
  perform highly undetectable steganography,'' in \emph{Information Hiding -
  12th International Conference, {IH} 2010, Calgary, {AB}, Canada, June 28-30,
  2010, Revised Selected Papers}, ser. Lecture Notes in Computer Science,
  R.~B{\"o}hme, P.~W.~L. Fong, and R.~Safavi-Naini, Eds., vol. 6387.\hskip 1em
  plus 0.5em minus 0.4em\relax Springer, 2010, pp. 161--177. [Online].
  Available: \url{http://dx.doi.org/10.1007/978-3-642-16435-4}
\BIBentrySTDinterwordspacing

\bibitem{FK2013}
J.~Fridrich and J.~Kodovský, ``Multivariate gaussian model for designing
  additive distortion for steganography,'' in \emph{Acoustics, Speech and
  Signal Processing (ICASSP), 2013 IEEE International Conference on}, May 2013,
  pp. 2949--2953.

\bibitem{DBLP:journals/tifs/HolubF15}
\BIBentryALTinterwordspacing
V.~Holub and J.~J. Fridrich, ``Low-complexity features for {JPEG} steganalysis
  using undecimated {DCT},'' \emph{{IEEE} Trans. Information Forensics and
  Security}, vol.~10, no.~2, pp. 219--228, 2015. [Online]. Available:
  \url{http://dx.doi.org/10.1109/TIFS.2014.2364918}
\BIBentrySTDinterwordspacing

\bibitem{DBLP:journals/tifs/KodovskyFH12}
\BIBentryALTinterwordspacing
J.~Kodovsk{\'{y}}, J.~J. Fridrich, and V.~Holub, ``Ensemble classifiers for
  steganalysis of digital media,'' \emph{{IEEE} Transactions on Information
  Forensics and Security}, vol.~7, no.~2, pp. 432--444, 2012. [Online].
  Available: \url{http://dx.doi.org/10.1109/TIFS.2011.2175919}
\BIBentrySTDinterwordspacing

\bibitem{DBLP:journals/tifs/HolubF13}
\BIBentryALTinterwordspacing
V.~Holub and J.~J. Fridrich, ``Random projections of residuals for digital
  image steganalysis,'' \emph{{IEEE} Trans. Information Forensics and
  Security}, vol.~8, no.~12, pp. 1996--2006, 2013. [Online]. Available:
  \url{http://dx.doi.org/10.1109/TIFS.2013.2286682}
\BIBentrySTDinterwordspacing

\bibitem{conf/mmsys/Dang-NguyenPCB15}
\BIBentryALTinterwordspacing
D.-T. Dang-Nguyen, C.~Pasquini, V.~Conotter, and G.~Boato, ``{RAISE}: a raw
  images dataset for digital image forensics,'' in \emph{Proceedings of the 6th
  {ACM} Multimedia Systems Conference, {MMS}ys 2015, Portland, {OR}, {USA},
  March 18-20, 2015}, W.~T. Ooi, W.~chi Feng, and F.~Liu, Eds.\hskip 1em plus
  0.5em minus 0.4em\relax ACM, 2015, pp. 219--224. [Online]. Available:
  \url{http://dl.acm.org/citation.cfm?id=2713168}
\BIBentrySTDinterwordspacing

\bibitem{lecun2015deep}
Y.~LeCun, Y.~Bengio, and G.~Hinton, ``Deep learning,'' \emph{Nature}, vol. 521,
  no. 7553, pp. 436--444, 2015.

\bibitem{krizhevsky2012imagenet}
A.~Krizhevsky, I.~Sutskever, and G.~E. Hinton, ``Imagenet classification with
  deep convolutional neural networks,'' in \emph{Advances in neural information
  processing systems}, 2012, pp. 1097--1105.

\bibitem{wan2013regularization}
L.~Wan, M.~Zeiler, S.~Zhang, Y.~L. Cun, and R.~Fergus, ``Regularization of
  neural networks using dropconnect,'' in \emph{Proceedings of the 30th
  International Conference on Machine Learning (ICML-13)}, 2013, pp.
  1058--1066.

\bibitem{2015arXiv151107289C}
D.-A. {Clevert}, T.~{Unterthiner}, and S.~{Hochreiter}, ``{Fast and Accurate
  Deep Network Learning by Exponential Linear Units (ELUs)},'' \emph{ArXiv
  e-prints}, Nov. 2015.

\bibitem{xu2016structural}
G.~Xu, H.-Z. Wu, and Y.-Q. Shi, ``Structural design of convolutional neural
  networks for steganalysis,'' \emph{IEEE Signal Processing Letters}, vol.~23,
  no.~5, pp. 708--712, 2016.

\bibitem{tan2014stacked}
S.~Tan and B.~Li, ``Stacked convolutional auto-encoders for steganalysis of
  digital images,'' in \emph{Asia-Pacific Signal and Information Processing
  Association, 2014 Annual Summit and Conference (APSIPA)}.\hskip 1em plus
  0.5em minus 0.4em\relax IEEE, 2014, pp. 1--4.

\bibitem{ker2008revisiting}
A.~D. Ker and R.~B{\"o}hme, ``Revisiting weighted stego-image steganalysis,''
  in \emph{Electronic Imaging 2008}.\hskip 1em plus 0.5em minus 0.4em\relax
  International Society for Optics and Photonics, 2008, pp. 681\,905--681\,905.

\bibitem{DBLP:journals/tifs/FridrichK12}
\BIBentryALTinterwordspacing
J.~J. Fridrich and J.~Kodovsk{\'{y}}, ``Rich models for steganalysis of digital
  images,'' \emph{{IEEE} Trans. Information Forensics and Security}, vol.~7,
  no.~3, pp. 868--882, 2012. [Online]. Available:
  \url{http://dx.doi.org/10.1109/TIFS.2012.2190402}
\BIBentrySTDinterwordspacing

\bibitem{qian2015deep}
Y.~Qian, J.~Dong, W.~Wang, and T.~Tan, ``Deep learning for steganalysis via
  convolutional neural networks,'' in \emph{IS\&T/SPIE Electronic
  Imaging}.\hskip 1em plus 0.5em minus 0.4em\relax International Society for
  Optics and Photonics, 2015, pp. 94\,090J--94\,090J.

\bibitem{pibre2016deep}
L.~Pibre, P.~J{\'e}r{\^o}me, D.~Ienco, and M.~Chaumont, ``Deep learning is a
  good steganalysis tool when embedding key is reused for different images,
  even if there is a cover source-mismatch,'' in \emph{Media Watermarking,
  Security, and Forensics, EI: Electronic Imaging}, 2016.

\bibitem{2016arXiv160507946C}
J.-F. {Couchot}, R.~{Couturier}, C.~{Guyeux}, and M.~{Salomon}, ``{Steganalysis
  via a Convolutional Neural Network using Large Convolution Filters for
  Embedding Process with Same Stego Key},'' \emph{ArXiv e-prints}, May 2016.

\bibitem{xu2016ensemble}
G.~Xu, H.-Z. Wu, and Y.-Q. Shi, ``Ensemble of cnns for steganalysis: An
  empirical study,'' in \emph{ACM Workshop on Information Hiding and Multimedia
  Security}, 2016.

\bibitem{tang2016adaptive}
W.~Tang, H.~Li, W.~Luo, and J.~Huang, ``Adaptive steganalysis based on
  embedding probabilities of pixels,'' \emph{IEEE Transactions on Information
  Forensics and Security}, vol.~11, no.~4, pp. 734--745, 2016.

\bibitem{denemark2016improving}
T.~Denemark, J.~Fridrich, and P.~Comesa{\~n}a-Alfaro, ``Improving
  selection-channel-aware steganalysis features,'' \emph{Electronic Imaging},
  vol. 2016, no.~8, pp. 1--8, 2016.

\bibitem{7532860}
Y.~Qian, J.~Dong, W.~Wang, and T.~Tan, ``Learning and transferring
  representations for image steganalysis using convolutional neural network,''
  in \emph{2016 IEEE International Conference on Image Processing (ICIP)}, Sept
  2016, pp. 2752--2756.

\bibitem{2016arXiv161103233Z}
J.~{Zeng}, S.~{Tan}, B.~{Li}, and J.~{Huang}, ``{Large-scale JPEG steganalysis
  using hybrid deep-learning framework},'' \emph{ArXiv e-prints}, Nov. 2016.

\bibitem{ioffe2015batch}
S.~Ioffe and C.~Szegedy, ``Batch normalization: Accelerating deep network
  training by reducing internal covariate shift,'' \emph{arXiv preprint
  arXiv:1502.03167}, 2015.

\bibitem{2016arXiv160304467A}
M.~{Abadi}, A.~{Agarwal}, P.~{Barham}, E.~{Brevdo}, Z.~{Chen}, C.~{Citro},
  G.~S. {Corrado}, A.~{Davis}, J.~{Dean}, M.~{Devin}, S.~{Ghemawat},
  I.~{Goodfellow}, A.~{Harp}, G.~{Irving}, M.~{Isard}, Y.~{Jia},
  R.~{Jozefowicz}, L.~{Kaiser}, M.~{Kudlur}, J.~{Levenberg}, D.~{Mane},
  R.~{Monga}, S.~{Moore}, D.~{Murray}, C.~{Olah}, M.~{Schuster}, J.~{Shlens},
  B.~{Steiner}, I.~{Sutskever}, K.~{Talwar}, P.~{Tucker}, V.~{Vanhoucke},
  V.~{Vasudevan}, F.~{Viegas}, O.~{Vinyals}, P.~{Warden}, M.~{Wattenberg},
  M.~{Wicke}, Y.~{Yu}, and X.~{Zheng}, ``{TensorFlow: Large-Scale Machine
  Learning on Heterogeneous Distributed Systems},'' \emph{ArXiv e-prints}, Mar.
  2016.

\bibitem{collobert2011torch7}
R.~Collobert, K.~Kavukcuoglu, and C.~Farabet, ``Torch7: A matlab-like
  environment for machine learning,'' in \emph{BigLearn, NIPS Workshop}, no.
  EPFL-CONF-192376, 2011.

\bibitem{Bas2011}
P.~Bas, T.~Filler, and T.~Pevn{\'y}, \emph{''Break Our Steganographic System'':
  The Ins and Outs of Organizing BOSS}.\hskip 1em plus 0.5em minus 0.4em\relax
  Berlin, Heidelberg: Springer Berlin Heidelberg, 2011, pp. 59--70.

\bibitem{DBLP:conf/wifs/DenemarkSHCF14}
\BIBentryALTinterwordspacing
T.~Denemark, V.~Sedighi, V.~Holub, R.~Cogranne, and J.~J. Fridrich,
  ``Selection-channel-aware rich model for steganalysis of digital images,'' in
  \emph{2014 {IEEE} International Workshop on Information Forensics and
  Security, {WIFS} 2014, Atlanta, GA, USA, December 3-5, 2014}.\hskip 1em plus
  0.5em minus 0.4em\relax {IEEE}, 2014, pp. 48--53. [Online]. Available:
  \url{http://dx.doi.org/10.1109/WIFS.2014.7084302}
\BIBentrySTDinterwordspacing

\end{thebibliography}
\end{document}